\documentclass[preprint]{elsarticle}
\usepackage{multirow}
\usepackage{color}
\usepackage{amsmath}
\usepackage{amsfonts}
\usepackage{amssymb}
\usepackage{subfigure}

\usepackage{graphicx}
\usepackage{balance}
\usepackage{colortbl}
\usepackage{multicol}
\usepackage{tabularx}
\usepackage{mathdots}
\usepackage{colordvi}
\usepackage{soul}

\usepackage{float}
\usepackage{hyphenat}

\usepackage{lineno,hyperref}
\modulolinenumbers[5]

\definecolor{orange}{rgb}{1,0.5,0}

\newcommand{\wam}[1]{\textcolor{black}{#1}}

\journal{EUSIPCO-2019}

\begin{document}

\begin{frontmatter}

\title{Intersymbol and Intercarrier Interference in OFDM Transmissions through Highly Dispersive Channels\tnoteref{thanks}}
\tnotetext[thanks]{This is a preliminary version of a paper to be submitted to EUSIPCO-2019. Personal use of this material is permitted. However, permission to use this material for any other purposes must be obtained from the authors by sending a request to the corresponding author.\\ This study was financed in part by the Coordena\c c\~ao de Aperfei\c coamento de Pessoal de N\'ivel Superior - Brasil (CAPES) --- Finance Code 23038.009440/2012-42. This work was partially supported by the Spanish Ministry of Economy and Competitiveness through project TEC2015-64835-C3-1-R.}

\author[ufrj]{Wallace Alves Martins\corref{mycorrespondingauthor}}
\cortext[mycorrespondingauthor]{Corresponding author}
\ead{wallace.martins@smt.ufrj.br}

\author[tsc]{Fernando~Cruz{--}Rold{\'a}n}
\author[kul]{Marc~Moonen}
\author[ufrj]{Paulo Sergio Ramirez Diniz}

\address[ufrj]{Electrical Engineering Program (PEE/Coppe) and Department of Electronics and Computer Engineering (DEL/Poli), Federal University of Rio de Janeiro (UFRJ), 21941-972, Rio de Janeiro/RJ, BRAZIL.}

\address[tsc]{Department of Signal Theory and Communications (EPS), Universidad de Alcal{\'a}, 28805 Alcal{\'a} de Henares (Madrid), SPAIN.}

\address[kul]{Department of Electrical Engineering (ESAT-STADIUS), KU Leuven, 3001 Leuven, BELGIUM.}

\fnref{a}

\begin{abstract}
This work quantifies, for the first time, intersymbol and intercarrier interferences induced by very dispersive channels in OFDM systems. The resulting  achievable data rate for \wam{suboptimal} OFDM transmissions is derived based on the computation of signal-to-interference-plus-noise ratio for arbitrary length finite duration channel impulse responses. Simulation results point to significant differences between data rates obtained via conventional formulations, for which interferences are supposed to be limited to two or three blocks, versus the data rates considering the actual channel dispersion.
\end{abstract}


\end{frontmatter}

\pagebreak


\section{Introduction}
Since its origins in the analog~\cite{Chang1966} and digital~\cite{Weins1971} domains, the orthogonal frequency-division multiplexing (OFDM) system has striven to combat interferences induced by frequency-selective channels~\cite{Weins2009}. A major breakthrough was the use of redundant elements, such as cyclic prefix, for preserving orthogonality among subcarriers at the receiver~\cite{Peled1980}. 

Practical wireless broadband communications use  increasingly high sampling rates while trying to meet the demands for high data rates. {In this context,} the redundancy overhead {turns out} to be a big issue due to spectral-efficiency losses.  Furthermore, as some practical wired applications  (e.g.,  digital subscriber lines, including ADSL, VDSL, and more recently, G.Fast) work with highly dispersive channels, it can be virtually impossible to append so many redundant elements in the transmission, calling for alternative solutions, such as prefiltering at the receiver side to shorten the effective channel model~\cite{Altin2011}.

The aforementioned issues motivated many works to analyze ISI/ICI in OFDM systems with insufficient redundancy~\cite{Viter1995,Seoan1997,Kim1998,Barh2004,Hen02,Milos2002,Vanb2004,
Zochm2014,Pham2017,Taher2016,Wang2017,Lim2017,Ouzzi2009,Wolke2013,
Barro2008,Zhong2006,
Monto2009}. 
Perhaps, the first attempt to analyze systematically the harmful ISI/ICI effects are the works in~\cite{Viter1995,Seoan1997, Pol97} in which preliminary results in terms of interference spectral power are presented. The authors in~\cite{Kim1998} analyze ISI interference in OFDM-based high-definition television (HDTV). The works in~\cite{Barh2004} address time-variant channels that induce Doppler effects. A landmark work is~\cite{Hen02}, which presents the interference power due to the tails of the channel impulse response (CIR). 
The works~\cite{Milos2002,Vanb2004
} use the ISI/ICI interference analyses to design the front-end prefilter employed at the receiver side for channel shortening in xDSL applications. 
The work in~\cite{Zochm2014} uses the results from~\cite{Hen02} to show it might be  useful to allow the existence of ISI/ICI due to insufficient guard periods, as long as the transceiver is aware of it---via feedback mechanism of channel-state information (CSI). 
The authors in~\cite{Ouzzi2009} analyze the ISI/ICI impacts on a power-line communication (PLC) system, whereas~\cite{Wolke2013
} optimize the redundancy length for a PLC system, allowing for the existence of controlled ISI/ICI. A similar kind of optimization is also conducted for coherent optical communications in~\cite{Barro2008}. 
The works~\cite{Zhong2006
} analyze the ISI/ICI effects from a theoretical and an experimental viewpoints for wireless local area networks. 
\wam{The works ~\cite{Schniter2007,Farhang2009,Schniter2004,Pecile2008} address the design/analysis of more general transceivers or equalizers, mostly for doubly-selective channel models.}

All of these works start from a common point---the analysis of ISI/ICI for a general setup where the amount of redundant elements can be smaller than the delay spread of the channel---to eventually achieve different goals. We shall start from the same point, but {\it without imposing any kind of constraints upon the order of the CIR}, which is only assumed to have finite duration. This eventually implies that the proposed analysis can be applied to the cases where the interference is due to several data blocks, not being limited to two or three blocks, as all previous works do. As mentioned before, this is especially important in delay-constrained applications working in highly dispersive environments where the number of subcarriers cannot be set larger than the order of the CIR. Besides, the proposed analysis unifies and generalizes many of the aforementioned works, providing closed-form \wam{matrix} expressions \wam{as functions of the channel taps} that are easy to use for different purposes---including the ones listed above. \wam{Both cyclic-prefix and zero-padding OFDM transmissions are considered and a fine distinction between two types of ICI is provided.} As for the application of the proposed analysis, this work focuses on providing {\it achievable data rates for \wam{suboptimal} OFDM transmissions} based on the computation of signal-to-interference-plus-noise ratios (SINRs) for arbitrary length finite duration CIR.

\section{System Model}\label{sec:ofdm}
\begin{figure}[t!]
\includegraphics[width=\linewidth]{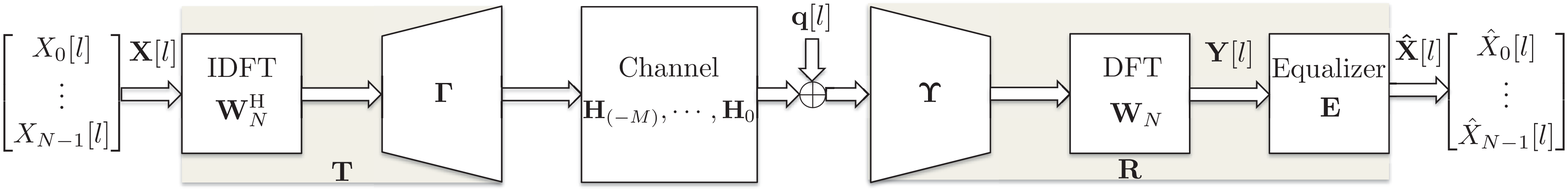}
\caption{Block diagram of an OFDM transceiver.}\label{fig:block_diag}
\end{figure}

Let us consider an OFDM transceiver model as depicted in  
Figure~\ref{fig:block_diag}. The data samples $X_k[l]$, with $k\in {\cal N} \triangleq \{0, 1, \cdots, N-1\}\subset\mathbb{N}$, belong to a particular constellation 
$ {\cal C} \subset \mathbb{C}$, such as QAM or 
PSK, and comprise vector ${\bf X}[l]$ at the (block) time index $l\in\mathbb{Z}$. After the transmission/reception process, the reconstructed data samples are denoted as $\hat{X}_k[l]$, with $k\in{\cal N}$, and comprise the  reconstructed data vector ${\bf \hat{X}}[l]$.

The cyclic-prefix OFDM (CP-OFDM) is described by the
following transmitter and receiver matrices, 
 respectively: 
\begin{align}
 \mathbf{T}_{\rm CP} &\triangleq \underbrace{\begin{bmatrix} \mathbf{0}_{\mu\times (N-\mu)}\quad {\bf I}_\mu \\ {\bf I}_N \end{bmatrix}}_{{\bf \Gamma}_{\rm CP} \in  \mathbb{C}^{N_0 \times N}}\cdot\mathbf{W}_N^{\rm H} \,,\\
\mathbf{R}_{\rm CP} &\triangleq {\bf E}\cdot\mathbf{W}_N\cdot\underbrace{\begin{bmatrix} \mathbf{0}_{N\times \mu} & \mathbf{I}_N\end{bmatrix}}_{{\bf \Upsilon}_{\rm CP} \in  \mathbb{C}^{N \times N_0}}\,,
\end{align}
where $N_0 \triangleq N + \mu \in\mathbb{N}$ denotes the size of the transmitted data vector after appending $\mu < N$ redundant elements,\footnote{In principle, there is no point in using $\mu \geq N$ in practice.} $\mathbf{W}_N$ is the normalized $N\times N$ discrete Fourier transform (DFT) matrix with entries $[\mathbf{W}_N]_{kn} = {\rm e}^{-{\rm j}\frac{2\pi}{N}kn}/\sqrt{N}$,  
${\bf I}_{N}$ is the $N\times N$ identity matrix, $\mathbf{0}_{N\times \mu}$ is an $N\times \mu$ matrix whose entries are zero, 
and $\mathbf{E} \in \mathbb{C}^{N \times N}$ is an equalizer  diagonal 
matrix. 

An alternative OFDM system inserts zeros as redundancy and 
is called  zero-padding OFDM (ZP-OFDM). 
There are many variants of ZP-OFDM. A common choice 
is the ZP-OFDM-OLA (overlap-and-add), with:
\begin{align}
\mathbf{T}_{\rm ZP} &\triangleq \underbrace{\begin{bmatrix} {\bf I}_N \\ \mathbf{0}_{\mu\times N}\end{bmatrix}}_{{\bf \Gamma}_{\rm ZP} \in  \mathbb{C}^{N_0 \times N}}\cdot\mathbf{W}_N^{\rm H} \,,\\
\mathbf{R}_{\rm ZP} &\triangleq 
{\bf E}\cdot\mathbf{W}_N\cdot
\underbrace{\left[\begin{tabular}{c c}
\multirow{2}{4mm}{$\mathbf{I}_N$} & $\mathbf{I}_\mu$ \\ & $\mathbf{0}_{(N-\mu)\times \mu}$  
\end{tabular}\right]}_{{\bf \Upsilon}_{\rm ZP} \in  \mathbb{C}^{N \times N_0}}\,.
\end{align}

The channel model is represented by a causal finite duration impulse response (FIR)  filter with coefficients $h_0, \ldots, h_\nu \in \mathbb{C}$ of order 
 $\nu\in \mathbb{N}$, and an additive noise vector  
 ${\bf q}[l] \in\mathbb{C}^{N_0}$. 
 In fact, the FIR model can also be regarded as an overall channel impulse response, encompassing any pulse shaping and front-end receive prefiltering. 
 Thus, assuming a synchronization delay $\Delta \in \{0, 1, \ldots , N_0 - 1\}$,\footnote{In fact, $\Delta$ could be any natural number, but we assume $\Delta \in \{0, 1, \ldots , N_0 - 1\}$ for the sake of simplicity.} the number of data vectors that may affect the reception of a single data vector is $M + 2$, where
\begin{equation}
M \triangleq \left\lceil\frac{\nu}{N_0}\right\rceil,
\label{eq:M-def}
\end{equation}
in which $\left\lceil \cdot \right\rceil $ stands for the ceiling function. Hence, the reconstructed data vector can be written as
\begin{align}
 {\bf{\hat X}}[l] = \sum_{m=-1}^M{\bf R}\cdot{\bf H}_{(-m)}\cdot{\bf T}\cdot{\bf X}[l-m] + {\bf R}\cdot{\bf q}[l], 
  \label{eq:estimated-vector}
\end{align}
where ${\bf{H}}_{(-m)}$ is an $N_0 \times N_0$ matrix, in which, for $0 \leq b,c \leq N_0-1$, one has
\begin{equation}
\left[ {{\bf{H}}_{( - m)} } \right]_{b,c}  \buildrel \Delta \over = \left\{ {\begin{array}{*{20}c}
   {0,} & {mN_0  + b - c + \Delta < 0,}  \\
   {h_{mN_0  + b - c+ \Delta} ,} & {0 \le mN_0  + b - c + \Delta \le \nu ,}  \\
   {0,} & {mN_0  + b - c + \Delta > \nu .}  \\
\end{array}} \right.
\label{eq:H-def}
\end{equation}

Note that in~\eqref{eq:estimated-vector}, up to $M+2$ data vectors may contribute to the reconstructed vector ${\bf{\hat X}}[l]$. In fact, this number can be smaller, depending on the delay $\Delta$. Indeed, based on~\eqref{eq:M-def}, one can write $\nu = (M-1) N_0 + \rho + 1$, with $\rho$ being a number in the set $\{0, 1, \ldots , N_0 - 1\}$. Based on~\eqref{eq:H-def}, if there is perfect synchronization (i.e., $\Delta = 0$), then ${\bf H}_{(1)} = {\bf 0}_{N_0 \times N_0}$ and the reconstructed data vector is affected by at most $M+1$ transmitted data vectors. In this case, the last matrix ${\bf H}_{(-M)}$ has only $\rho + 1$ nonzero rows. This eventually implies that, for $\Delta > \rho$, what would be the $(M+2)^{\rm th}$ channel matrix will actually be a null matrix, and again up to $M+1$ vectors affect ${\bf{\hat X}}[l]$. In summary, the number of data vectors contributing to the reconstructed vector is up to $M+2$ whenever $\Delta$ is chosen within the set $\{1, 2, \ldots , \rho\}$, or otherwise up to $M+1$.

Considering the scenarios in which the redundancy is long enough, i.e. $1 \leq \nu\leq \mu$, one has $M = 1$. In that case, the received {data vector} at time $l$ might be affected only by the transmitted data vectors at times $l+1$, $l$, and $l-1$, for a nonzero delay $\Delta$. As mentioned before, when $\Delta=0$, one has ${\bf H}_{(1)} = {\bf 0}_{N_0 \times N_0}$, and 
matrices  $\mathbf{\Upsilon}$ and $\mathbf{\Gamma}$ are able to eliminate the interference induced by the transmitted data vector  at $l-1$, i.e., ${\bf \Upsilon}\cdot{\bf H}_{(-1)}\cdot{\bf \Gamma} = {\bf 0}_{N \times N}$. In addition, one has that ${\bf \Upsilon} \cdot {\bf H}_{0} \cdot {\bf \Gamma}$ is a right-circulant matrix of dimension $N\times N$ that can be diagonalized by the DFT matrix. As a result,  
\begin{align}\label{eq:estimated-vector-case1}
{\bf \hat{X}}[l] = {\bf E}\cdot{\bf D}\cdot{\bf X}[l] + {\bf R}\cdot{\bf q}[l], 
\end{align}
where the diagonal matrix ${\bf D}$ is 
\begin{align}\label{eq:D}
{\bf D} 
&\triangleq {\rm diag}\left\{\sqrt{N}{\bf W}_N\cdot\begin{bmatrix}{\bf h}\\{\bf 0}_{(N-\nu-1)\times 1}\end{bmatrix}\right\},
\end{align}
in which ${\bf h} \triangleq [\,h_0\;\;h_1\;\;\cdots\;\;h_\nu\,]^{\rm T}$. 
As can be noted, there are no ISI or ICI when $\nu \leq \mu$.

The equalizer ${\bf E}$ for this transceiver can be defined in several ways,
where the most popular ones are 
the zero-forcing (ZF) and the minimum mean-squared error (MMSE) equalizers, with 
${\bf E}_{\rm ZF} \triangleq {\bf D}^{-1}$ or 
${\bf E}_{\rm MMSE} \triangleq {\bf D}^H\cdot\left({\bf D}\cdot{\bf D}^H + \frac{1}{\rm SNR}{\bf I}_N\right)^{-1}$, 
where SNR stands for signal-to-noise ratio. In the latter case, the transmitted symbols and
environment noise are  wide-sense stationary (WSS), mutually independent, white random sequences. 

As mentioned before, some previous works have analyzed the case where $\nu > \mu$, but with the restriction of having $\nu \leq N_0$. Next section presents an analysis for general $\nu$.

\section{ISI/ICI Analysis}\label{sec:analysis}
This section focuses on OFDM transceivers with insufficient number of redundant elements, i.e., $\nu > \mu$. In this case, the received data vector at time $l$ can be affected by the transmitted data vectors at times $l+1, l, l-1, \ldots, l-M$, and  matrices  $\mathbf{\Upsilon}$ and $\mathbf{\Gamma}$ cannot eliminate all ISI/ICI. 

One can rewrite eq.~\eqref{eq:estimated-vector} as
\begin{eqnarray}
{\bf{\hat X}}[l]  &=&  
  \sum_{\scriptstyle m = -1 \hfill \atop 
  \scriptstyle m \ne 0 \hfill}^{M}{\bf E}\cdot\underbrace {{\bf W}_N \cdot\mathbf{\Upsilon} \cdot \mathbf{H}_{(-m)}\cdot\mathbf{\Gamma}\cdot {\bf W}_N^{\rm H}}_{{\bf{A}}^{\rm ISI,ICI_2 }_{m}}\cdot\mathbf{X}[l-m]\nonumber \\
  & & + \;{\bf E}\cdot\underbrace {{\bf W}_N \cdot\mathbf{\Upsilon} \cdot \mathbf{H}_0\cdot\mathbf{\Gamma}\cdot {\bf W}_N^{\rm H}}_{{\bf{B}}^{\rm des,ICI_1 }}\cdot\mathbf{X}[l] + \;{\bf E}\cdot \underbrace {{\bf W}_N \cdot\mathbf{\Upsilon}}_{{\bf{G}}^{\rm noise} }\cdot \mathbf{q}[l].
\label{eq:Xr_isi_ici}
\end{eqnarray}

\noindent The desirable signal of~\eqref{eq:Xr_isi_ici} is
\begin{equation}
\mathbf{\hat{X}}_{\rm des}[l]=  {\bf E} \cdot {\mathbf{B}}^{\rm des} \cdot \mathbf{X}[l],
\label{eq:Xr_desM1}
\end{equation}
in which $\mathbf{B}^{\rm des}$ is a diagonal matrix with elements
\begin{equation}
\left[ \mathbf{B}^{\rm des} \right] _{i,i}\triangleq \left[ {\mathbf{B}}^{\rm des,ICI_1} \right]_{i,i}.
\end{equation}

The difference between \eqref{eq:Xr_isi_ici} and \eqref{eq:Xr_desM1} defines the ISI and ICI. Now, the products $\mathbf{\Upsilon} \cdot \mathbf{H}_{(-m)} \cdot \mathbf{\Gamma}$, cannot be expressed as $N \times N$ right-circulant matrices, and therefore they cannot be diagonalized using DFTs. Based on~\cite{Pol97}, the interference can be classified into three different types:
\begin{itemize}
\item {ISI: This} is the interference from the data vector transmitted at time $l- m$, with $m\in\{-1,1,2, \ldots, M\}$, in the considered data vector transmitted at time $l$ on the same subcarrier. All diagonal elements $\left[\mathbf{A}_m^{\rm ISI,ICI_2}\right]_{i,i}$ contribute to this interference. 
\item Type-1 ICI (ICI$_1$): This is the interference among different subcarriers belonging to the considered data vector transmitted at time $l$. It appears as a consequence of the elements $\left[\mathbf{B}^{\rm ICI_1}\right]_{i,j}$, $i \neq j$, where  $\mathbf{B}^{\rm ICI_1}\triangleq\mathbf{B}^{\rm des,ICI_1}-\mathbf{B}^{\rm des}$. Note that $\left[\mathbf{B}^{\rm ICI_1}\right]_{i,i} = 0$. 
\item {Type-2 ICI (ICI$_2$):  This} is the interference among different subcarriers of the data vector transmitted at time $l- m$, with $m\in\{-1,1,2 \ldots, M\}$, in the considered data vector transmitted at time $l$. The elements $\left[\mathbf{A}_m^{\rm ISI,ICI_2}\right]_{i,j}$, $i \neq j$, contribute to this interference. 
\end{itemize}
Finally, the contribution of noise $\mathbf{q}[l]$ to the reconstructed data vector $\mathbf{\hat X}[l]$ depends on matrix $\mathbf{G}^{\rm noise}$.

\section{\wam{SINR Analysis and Applications}}\label{sec:datarates}
This section presents an application of the previous ISI/ICI analysis for \wam{conducting an SINR analysis of} OFDM systems\wam{, which can then be employed in many applications, including computing achievable data rates}. We shall start with the signals before the multiplication by the equalizer matriz $\bf E$. Based on \eqref{eq:Xr_isi_ici} and~\eqref{eq:Xr_desM1}, one has
\begin{align}
{\bf Y}[l] \triangleq&\;
  {\bf{B}}^{\rm des,ICI_1 }\cdot\mathbf{X}[l]
    + \sum_{\scriptstyle m = -1 \hfill \atop 
  \scriptstyle m \ne 0 \hfill}^{M}{\bf{A}}^{\rm ISI,ICI_2 }_{m}\cdot\mathbf{X}[l-m] + {\bf{G}}^{\rm noise} \cdot \mathbf{q}[l],\\
\mathbf{Y}_{\rm des}[l] \triangleq & \; {\mathbf{B}}^{\rm des} \cdot \mathbf{X}[l].
\end{align}

Now, assuming the transmitted symbols and the noise signal are WSS, mutually independent, white random sequences with zero means and variances $\sigma_X^2$ and $\sigma_Q^2$, respectively, then 
one can compute the covariance matrices of the desired signal (${\bf C}_{\rm s}$), of the noise component (${\bf C}_{\rm n}$), and of ISI/ICI (${\bf C}_{\rm i}$). Indeed, one has
\begin{align}
{\bf C}_{\rm s} &= E\left\{\mathbf{Y}_{\rm des}[l]\cdot\mathbf{Y}_{\rm des}^{\rm H}[l]\right\} = \sigma_X^2 \cdot{\mathbf{B}}^{\rm des}\cdot ({\mathbf{B}}^{\rm des})^{\rm H}, \label{signal_power}\\
{\bf C}_{\rm i}+{\bf C}_{\rm n} &= E\left\{({\bf Y}[l]-\mathbf{Y}_{\rm des}[l])\cdot({\bf Y}[l]-\mathbf{Y}_{\rm des}[l])^{\rm H}\right\} \nonumber\\
&= \underbrace{\sigma_X^2\left({\mathbf{B}}^{\rm ICI_1}\!\cdot\! ({\mathbf{B}}^{\rm ICI_1})^{\rm H}+\sum_{\scriptstyle m = -1 \hfill \atop 
  \scriptstyle m \ne 0 \hfill}^M{\mathbf{A}}_{m}^{\rm ISI, ICI_2}\!\cdot\!({\mathbf{A}}_{m}^{\rm ISI, ICI_2})^{\rm H}\right)}_{{\bf C}_{\rm i}}\nonumber\\
&\quad+\underbrace{\sigma_Q^2 \cdot{\bf{G}}^{\rm noise}\cdot ({\bf{G}}^{\rm noise})^{\rm H}}_{{\bf C}_{\rm n}}. \label{ici_isi_power}
\end{align}

Hence, the SINR related to the $k^{\rm th}$ subcarrier is given as
\begin{align}\label{SINR}
{\rm SINR}(k) &= \frac{[{\bf C}_{\rm s}]_{kk}}{[{\bf C}_{\rm i}]_{kk}+[{\bf C}_{\rm n}]_{kk}}.
\end{align}

When QAM is used and error probability is measured in terms of symbol error rate (${\rm SER}$), the achievable data rate for  the $k^{\rm th}$ subcarrier is
\begin{equation}\label{eq_capacityDCT}
C \left( k \right)= \log _2 \left(1+ {\frac{{{\rm SINR}\left( k \right)}}{\Gamma }} \right),
\end{equation}
in which the SNR gap $\Gamma$ for a target ${\rm SER}$ is $\Gamma= \gamma_{\rm dm}-\gamma_{\rm c}+\Gamma_{\rm m}$ (in decibels), where $\gamma_{\rm dm}$ is a design margin, $\gamma_{\rm c}$ is the coding gain, and 
\begin{align}
\Gamma_{\rm m}  = \frac{1}{3}\left[ {Q^{ - 1} \left( {\frac{{{\rm SER}}}{4}} \right)} \right]^2 ,
\end{align}
with $Q(\cdot)$ being the tail distribution function of the normal distribution. 

\wam{By disregarding the signal correlation among the transceiver sub-channels, the} achievable data rate \wam{of a suboptimal system can finally be} obtained as
\begin{equation}\label{eq_throughputDCT}
R = f_s \cdot \frac{N}{{N_0}} \cdot \sum\limits_{k\in{\cal N}}{ C\left( k \right)},
\end{equation}
where $f_s$ is the underlying sampling rate.

\section{Simulation Results}\label{sec:sim}
The SINR calculation is an important task for several applications, including the design of time-domain equalizers (TEQs) and the calculation of achievable data rates. In this context, this section exemplifies two key points: the influence that the limitation of CIR length has on  TEQ design, and the differences of calculating the achievable data rate considering the interference limited to two or three blocks (Conventional), versus the one considering the actual dispersion (Actual).

\begin{figure}\centering
\includegraphics[width = 0.65\linewidth]{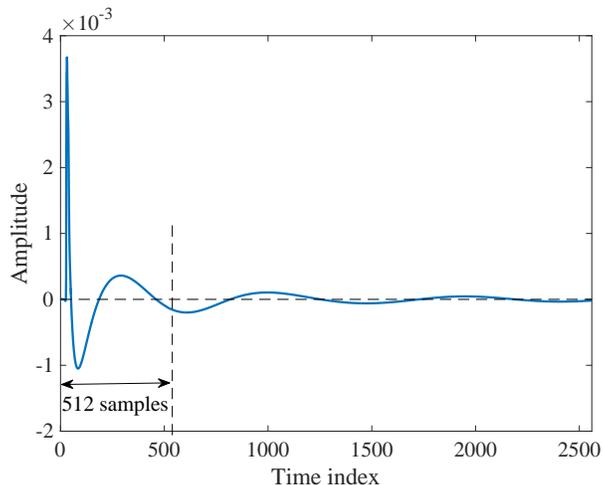}
\caption{Response of the CSA loop 4 in series with a high-pass filter.\label{fig:impulse-response-csa4}}
\end{figure}

A classic simulation setup, based on the widely used carrier-serving area (CSA) downstream loops~\cite{Milos2002,Vanb2004}, is employed to achieve this goal. In the experiments, the CSA loops are in series with a fifth-order Chebyshev Type-I high-pass filter with cut-on frequency at $4.8$~kHz, which filters out the telephone voiceband signal~\cite{Mart2005}. For downstream, the IDFT and DFT have size $N=512$, which is also the maximum CIR length in the Conventional analysis for obtaining the SINR. However, the convolution of any CSA loop with the high-pass filter has significant samples beyond the 512-sample  index. Figure~\ref{fig:impulse-response-csa4} depicts an example of the convolution of CSA loop 4 with the high-pass filter. The discarded samples beyond the time index $512$ correspond to $21.18 \%$ of the effective CIR energy. The Actual analysis via the proposed formulation considers all samples in Figure~\ref{fig:impulse-response-csa4}. 

To use relatively small CP lengths, one can employ TEQs, which can be designed via classic techniques, such as those proposed in~\cite{Melsa96} (MSSNR) and~\cite{Farhang01} (EIGAP). The design task is conducted considering: the first 512 samples in Figure~\ref{fig:impulse-response-csa4}  (Conventional), or the entire response in Figure~\ref{fig:impulse-response-csa4} (Actual). Once the TEQs have been designed, the overall impulse responses (OIRs) are obtained by convolving the response in Figure~\ref{fig:impulse-response-csa4} with the TEQ response. From these OIRs, the achievable data rates are obtained with the interference limited to two or three blocks (Conventional) or to a larger number of blocks (Actual). 

Although there are several sources of DSL noise, the achievable data rates are computed by considering only additive white Gaussian noise (AWGN) at -140 dBm/Hz, assuming an input signal power of $23$~dBm/Hz~\cite{Vanb2004}. All powers are defined with respect to a 100-$\Omega$ resistor. The following parameters were chosen to compute the data rate for each subcarrier $k$: an SNR gap $\Gamma_{\rm m}=9.8$~dB (for a ${\rm SER}=10^{-7}$), a noise margin $\gamma_{\rm dm} = 6$~dB, and a coding gain $\gamma_{\rm c}=4.2$~dB. The sampling rate is $f_s = 2.208$~MHz. The active tones are $\{7, 8, \ldots, 256\}$, and the TEQ time-offset has been optimized over the values $\{2, 3, \ldots, 50\}$.
 
\begin{figure}\centering
\subfigure[]{\includegraphics[width = 0.65\linewidth]{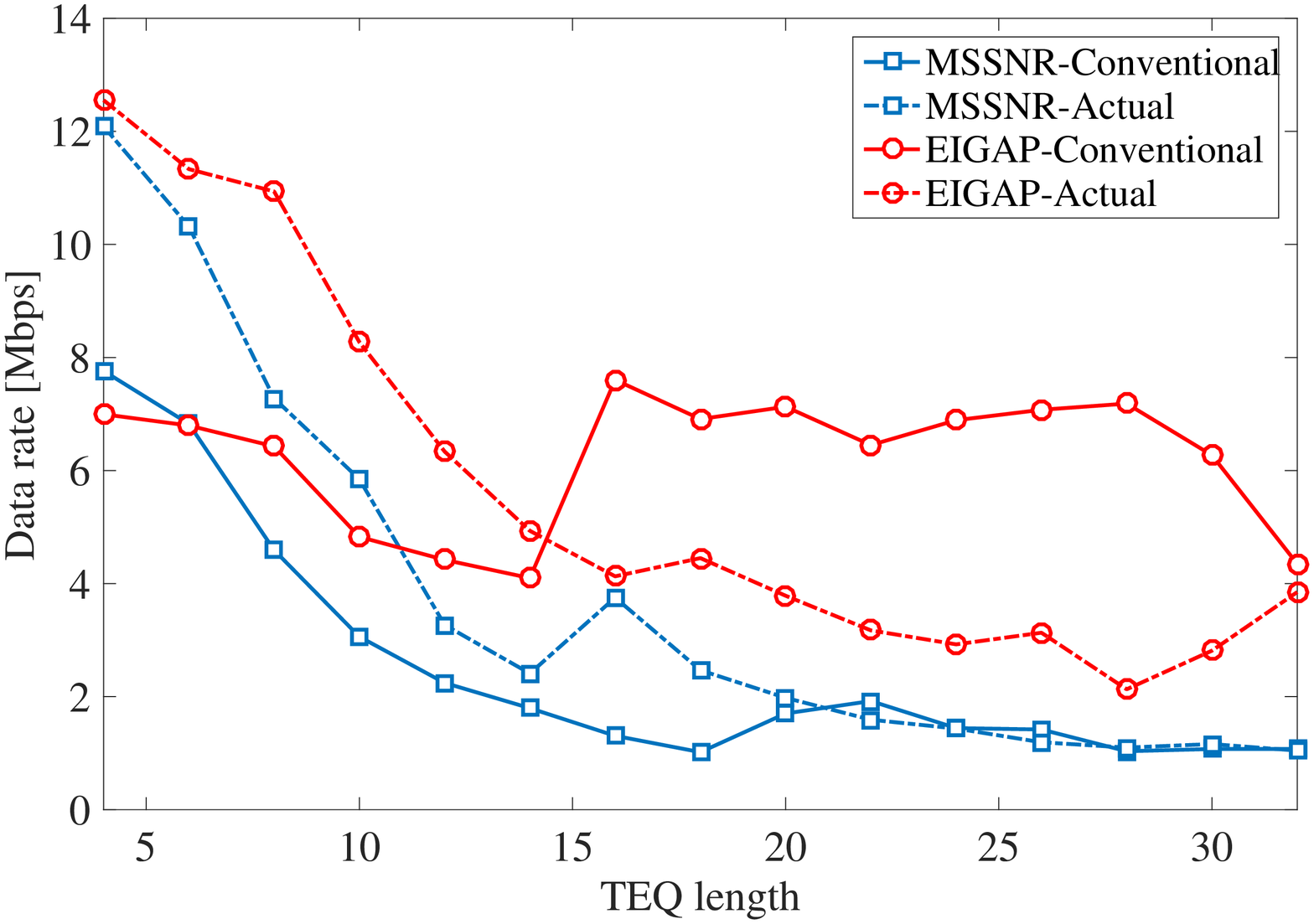}
\label{fig:Throughput_longiteq}}
\subfigure[]{\includegraphics[width = 0.65\linewidth]{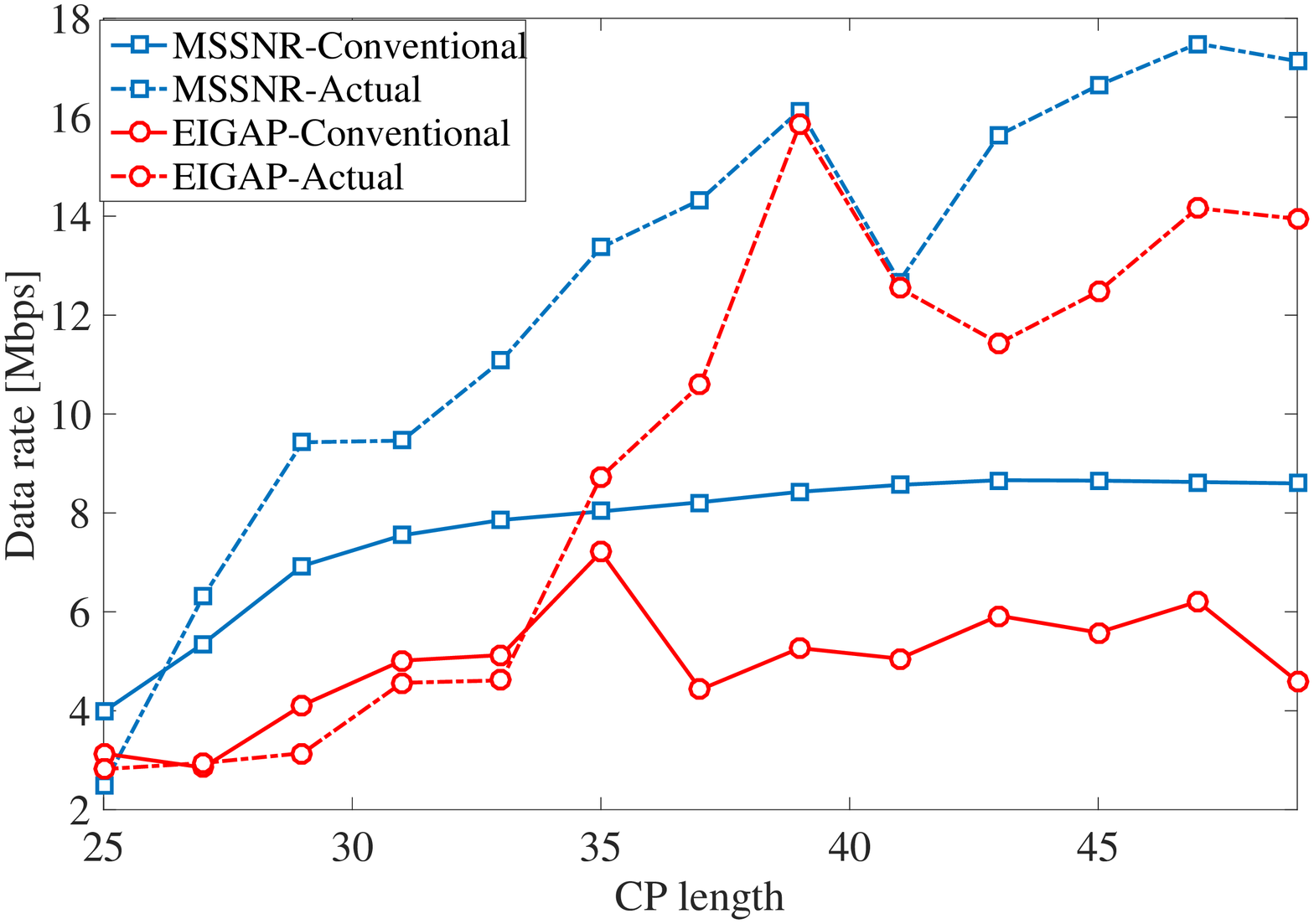}
\label{fig:Throughput_longiCP}}
\caption{Achievable data rates as a function of (a) TEQ and (b) CP lengths.}
\end{figure}

Figures~\ref{fig:Throughput_longiteq} and~\ref{fig:Throughput_longiCP} respectively depict the achievable data rates as a function of the TEQ lengths (for a fixed CP length of $\mu = 32$) and as a function of the CP length (for an optimized TEQ length in the range of values from 2 to the CP length of each experiment). The goal of these simulations is not comparing the design techniques of~\cite{Melsa96, Farhang01}, but showing the differences in the results of the TEQ designs or in the calculation of data rates. As can be seen in Figures~\ref{fig:Throughput_longiteq} and~\ref{fig:Throughput_longiCP}, the results obtained differ for virtually all the design parameters considered. These differences highlight the importance of using a correct formulation for the above calculations.

\section{Concluding Remarks}\label{sec:conc}
This work discussed the impact of highly dispersive channels on OFDM systems when the length of the prefix appended to the transmission block does not meet the requirement to induce uncoupled equalization solution. We derived the SINR and the achievable data rate for \wam{suboptimal} OFDM systems under arbitrary length finite duration CIR. In addition, and based on the analytical expressions, the work provided some simulations showing the differences between the results obtained by assuming that interference is limited to two or three blocks, versus those considering all interference blocks. In this sense, the theoretical expression derived for the SINR is more suitable to practical scenarios.

\section*{References}

\bibliography{fcruz-wam}

%
%
%
%
%

\end{document}